# Thoughts on the Proteins Native State

Jorge A. Vila

IMASL-CONICET, Universidad Nacional de San Luis, Ejército de Los Andes 950, 5700 San Luis, Argentina.

## Abstract

The presence of metamorphism in the protein's native state is not yet fully understood. In an attempt to throw light on this issue here we present an assessment, in terms of the amide hydrogen exchange protection factor, that aims to determine the likely existence of structural fluctuations in the native-state consistent with both the upper bound marginal stability of proteins and the metamorphism presence. The preliminary results enable us to conclude that the native-state metamorphism is, indeed, more probable than thought.

Let us start recalling the research that lead Anfinsen[1] to obtain the Nobel price "… *The work that my colleagues and I have carried out on the nature of the process that controls the folding of polypeptide chains into the unique three-dimensional structures of proteins was, indeed, strongly influenced by observations on the ribonuclease molecule…*" Now, what a "unique three-dimensional structure" for Anfinsen is? In his own words "*…is the one in which the Gibbs free energy of the whole system is lowest…*" There is no contradiction between Anfinsen's unique structure concept and the existence of an ensemble of folded conformations coexisting with the native state,[2-8] essentially, because all those conformations must possess higher Gibbs free energy, although comparable, to that of the native state.[9] From a statistical-mechanics point of view, this means that the native form will be the most probable form of the ensemble and any other conformation will have a representation consistent with their Boltzmann factor.

From the above analysis two basic questions arise. Firstly, which is the origin of the ensemble of folded conformations in equilibrium with the native state? Secondly, how large the structural difference among the conformations of the ensemble could be?

Regarding the first question, it is well known that the native state is marginally stable[3,4,10] and, hence, stabilized by weak interactions, such as those due to the interplay of pairwise and many-body interactions on both the proteins and the solvent;[11-13,8] therefore, one may expect transient fluctuations in the native structure to occur at a sizeable level. This is a crucial issue since it contains the origin of the protein native-state metamorphism,[9,14-17] which is a highly important phenomenon in structural biology[18] and protein evolvability.[7,15,19-21] Illustrations of this phenomenon are abundant and some of them have been recently reviewed in light of Anfinsen's dogma;[22] indeed, we have been able to show that the coexistence among highly-similar folded states, *e.g.*, for ubiquitin,[6] α–chymotrypsin, ribonuclease A, cytochrome *c*, etc.,[23] or between highly-dissimilar folded states, *e.g.*, the spindle checkpoint protein Mad-2[14] and the human chemokine lymphotactin (Ltn),[16] did take place and satisfied the thermodynamic hypothesis.[1]

An explanation to the second question, stated above, is not so straightforward and, hence, it will be investigated below in terms of the amide hydrogen exchange protection factor.[24,25] Before we embark on elaborate a possible solution to this problem, let us remember that proteins that fulfill the Anfinsen dogma possess an upper bound marginal stability, namely $\Delta G \leq \sim 7.4$ kcal/mol.[8,9] This implies that the largest free-energy difference among any conformation of the ensemble of folded states and the native-state (see Figure 1) must be lower than such upper bound, otherwise, the conformation will unfold, *i.e.*, will become nonfunctional.[8] As to whether such upper bound protein marginal stability limit (ΔG ~7.4 Kcal/mol) is large enough to assure the existence of significant topological differences in the ensemble of conformations, *i.e.*, to account for the existence of proteins native state metamorphisms, will be here examined in terms of the amide Hydrogen eXchange (HX) rate which is a sensitive probe to the stability of proteins.[23-27]

The protein marginal stability upper bound plays a central role in our analysis because it determines a threshold under which a protein kept folded and functioning. This means that we should begin by reexamining its physical origin and the theory that underlies its determination, which includes, but is not limited to, the use of (*i*) basic relationships derived from statistical thermodynamics, *e.g.* the Gibbs (G) free energy of the system can be obtained, in the thermodynamic limit, by the maximum eigenvalue of the transfer matrix,[28] whose elements are Boltzmann factors; (*ii*) the Gershgorin (circle) theorem[29] -that enabled us to determine an '*upper bound*' for the maximum eigenvalue of the transfer matrix and, consequently, for the Gibbs free energy; (*iii*) the assumption that the largest free-energy difference (ΔG) among coexisting native folds must verify two conditions: firstly, ΔG > 1, because the native-state is the conformation, or conformations if this state is degenerated, for which the Gibbs free energy of the whole system is lowest, and secondly, ΔG should be small, because all native folds in equilibrium with the native-state must possess higher, although comparable, free energy;[9] and (*iv*) a heuristic argument based on the main forces that stabilize proteins -that enabled us to find out a feasible analytical solution to a, otherwise, unsolvable rational fraction.[9] Should be noted that step (*iii*) of the theory, at different from the other steps, outlines the physical origin of the protein marginal stability. Indeed, it is stated as a necessary condition that the thermodynamic hypothesis -or Anfinsen dogma- must be fulfilled (ΔG > 1) assuring, in this way, that the native-state is '*stable*'. Moreover, it is required that ΔG be small, suggesting that the native-state stability is governed by weak interactions,[10-13] and, hence, should be '*marginal*'.[8] Altogether, the theoretical steps outlined above enabled us to find an analytical expression to forecast a robust upper bound for the protein's marginal stability, namely, as $\Delta G \leq \mathcal{L}im_{MW \to \infty} RT \ln MW$;[9] where *MW* stands for the protein molecular weight, *R* the gas constant and *T* the absolute temperature. The robustness of the forecast upper bound for ΔG (~7.4 Kcal/mol) arises from its logarithm dependence with *MW*, *i.e.*, that smooth the impact of possible differences with the assumed value -in the thermodynamic limit- for this parameter.[8] In sum, the marginal stability of proteins is essentially a consequence of the Anfinsen's dogma validity and its upper bound (~7.4 Kcal/mol), obtained regardless of the fold-class or sequence, a universal feature of proteins.

The use of hydrogen exchange for the purpose of determine protein stability is well known[23,25,30-32] since pioneer experiments of Linderstrøm-Lang and coworkers [33-36] and hence will not be here reviewed.

When a protein is transferred from water to deuterium there is a fast exchange of protons exposed to the solvent, namely an isotopic exchange. However, the backbone amide protons exchange will not take place if they are participating in intramolecular hydrogen bonding unless a transient opening in the native state of the protein occurs. Since both the intramolecular hydrogen bonding and the solvent shielding are straightforwardly related to the protein native state structure, it is reasonable to assume that the amide hydrogen exchange could be used as a sensitive probe to appraise changes in the protein topology.[23] Hence, adopting the kinetic scheme of Bahar *et al.*[24] in the EX2 limit, which describe the amide HX rate in the native state, the following relation should hold:

$$\Delta G_{HX} = -RT \ln K_{op} \qquad (1)$$

where $\Delta G_{HX}$ represent the Gibbs free-energy change for the opening/closing equilibrium;[24,27,31] the conversion of HX rates to Gibbs free-energies is possible because in the EX2 limit "…*the observed exchange rates are a measure of the equilibrium fraction of molecules that are available for exchange...*"[26], $R$ the gas constant, $T$ the absolute temperature, $K_{op} = 1/P_f$ is the equilibrium constant of the opening process[24,31] and $P_f$ the protection factor, which provides a measure of the resistance of the amide hydrogen to exchange in the native state relative to that of a random polypeptide chain.[24] However, our interest is focused on a particular region of the protein conformational space, namely, the ensemble of folded states in equilibrium with the native state (the yellow region in Figure 1). Therefore, the foresee value for the largest Gibbs free-energy change should be $\Delta G_{HX} \approx \Delta G$ (~7.4 kcal/mol). Under this conjecture, the protection factor ($P_f$) computed with Eq. (1) should represent the resistance of amide hydrogen to exchange in the native state relative to that of the highest free-energy conformation in the ensemble of folded states. Consequently, at room temperature, the amide hydrogen exchange protection factor upper bound, computed with Eq. (1), will be $P_f \sim 10^5$. In this regard, is worth noting that the computed upper bound for the protection factor (~$10^5$) does not necessarily reflect what can be observed at 'residue-level'. Indeed, the $β_3$ region of bovine pancreatic ribonuclease A[37] (RNase A) can reach $P_f$ 'peak values' of ~$10^5$; while, a much larger 'local $P_f$' can found in other proteins, such as for the $β_1$ region of bovine pancreatic trypsin inhibitor (BPTI) or the $α_5$ region of cytochrome *c*, where $P_f$'s can reach peak values of ~$10^8$ and ~$10^9$, respectively.[24] This contrast, in terms of the higher local $P_f$'s, illustrate that the 'local' protection factors can vary significantly along the amino-acid sequence, reflecting different tendency to undergo local structural changes. In particular, each of these highest local $P_f$ value belongs to buried, hydrogen-bonded, structures with a $\Delta G_{HX}$ value that, for some proteins, is comparable to that of the global unfolding free-energy ($\Delta G^U$), *e.g.*, the $β_3$ region of RNase A show, at 35 ºC, a $\Delta G_{HX} \cong \Delta G^U$ ~7.2kcal/mol.[37] From this point of view, a protection factor upper bound of ~$10^5$ enables us to conjecture that major structural change in the ensemble of native folds is highly likely. Overall, analysis of the 'local $P_f$' variations among proteins suggest that the local structural changes in the native state depends on many factors, among others, the local packing density, the topology of tertiary contacts, changes in the milieu, *etc.*, making the forecast of the extent of local structural changes, for a given sequence, very hard to determine accurately; a problem that will be exacerbated in the absence of a three-dimensional structure.

The main drawback of our analysis, so far, is that we fail to provide conclusive proof on the 'magnitude' of the local structural changes that could take place in the ensemble of folded states in equilibrium with the native state. The reason for this failure obeys, as explained above, the fact the largest local structural changes in native folds are not a universal feature of protein's but one depending, mainly, on the interactions of the residues with their nearest neighbors; in other words, it is determined by the fold-class and amino-acid sequence. This means that the only known universal feature of the native state of proteins is the existence of marginal stability upper bound ($\Delta G$ ~7.4 kcal/mol). Therefore, the computed amide HX protection factor upper bound (~$10^5$) should also be, by definition, a universal feature of proteins and, consequently, enable us to suggest the existence of an ensemble of folded states that may exhibit significant structural differences, relative to their native state, independent of protein's sequence or fold-class;[9] with the warning that the folded states ratio is determined by their Boltzmann factors. Then, how is that more than one native fold could be energetically favored?[17] A simple and feasible solution to this enigma is the following. Let us assume that small spontaneous changes in the milieu may indeed happens,[15,38] *i.e.*, on the solvent, pH, ionic strength, metal ions, temperature, etc. This could allow redistribution of the Boltzmann factors and, hence, a new thermodynamic equilibrium between highly-dissimilar (metamorphic) folded states would in fact be possible without challenging the Anfinsen dogma,[22] *e.g.*, the relative abundance of coexistent folded states on the aforementioned metamorphic proteins Mad2 and Ltn is, indeed, a thermodynamic-driven process.[14] It is worth noting that a switch between native-folds can also be trigger by mutations or cellular stimuli.[39]

The fact that amide hydrogen exchange is a powerful and versatile tool with which the dynamics of folded and unfolded states of proteins can be accurately determined has led to the development of a plethora of theoretical models and numerical simulations aimed to deeply understand the nature of the phenomenon.[25,40-43] Among all these studies, called our attention that one of them was able to show "… *that it is possible to obtain a detailed model for the structures populated during the rare equilibrium fluctuations of the native state that are sampled by hydrogen exchange*..."[40] The foretold structural fluctuations -up to 6Å root-mean-square-deviation, at the residue level, from the native state- are required for an accurate analysis of the amide hydrogen exchange, as evidenced by the attain good correlation between simulated and experimental protection factors, *viz.*, $R$ ~0.9.[40] What matters the most, those fluctuations conform to our conjecture that sizeable structural differences in the ensemble of folded states relative to the native state are indeed likely.

Overall, the take-home message of this work is twofold. Firstly, there is not a unique conformation representing protein's native states, *i.e.*, proteins are not monomorphic as is recurrently assume, but an ensemble of native-folds that coexist in a narrow range of free-energy variations determined by the physical origin of the marginal stability, which is a universal property of proteins.[8,9] Secondly, the aforementioned protein marginal stability is large enough to enable us suggest that native state metamorphism in proteins could be a highly probable phenomena; as inferred from the computed amide hydrogen exchange protection factor upper bound for the ensemble of folded states (~$10^5$). This conclusion, in addition, enables us rationalize the origin of the widespread observation, in the structural biology field, that "…*proteins that can adopt more than one native folded conformation may be more common than previously thought*…"[15]

7**Acknowledgments**

I am honored to dedicate this manuscript to the memory of Harold A. Scheraga, Professor of Chemistry, Emeritus at Cornell University. With the death of Harold Scheraga in Ithaca, NY, on August 1st, 2020, there passed one of the most distinguished figures of what may be called the heroic age of *ab initio* protein structure prediction; indeed, he was pioneer in the development of all-atoms force-fields, namely, the Empirical Conformational Energy Program for Peptides -ECEPP- introduced in 1975, aimed to predict the three-dimensional structure of proteins with no other information than the amino acid sequence. This, together with a long list of achievement, among others, the determination of the (*i*) structure of liquid water; (*ii*) oxidative folding pathways of RNase A; (*iii*) experimental parameters to characterize the helix-forming tendency of all 20 naturally occurring amino acids -with the host-guest technique-; (*iv*) crystal structures of small organic compounds -without knowledge-based information such as space groups, unit cell dimensions, or experimental powder diffraction patterns-; (*v*) accuracy of X-ray and NMR determined protein structures -based on quantum chemical calculations at the DFT-level of theory-, etc., enabled Harold to achieve leadership in the world of science, and high respect among colleagues.

The author acknowledges financial support from the IMASL-CONICET (PIP-0087) and ANPCyT (Grants PICT-0767; PICT-02212), Argentina.
Corresponding author
Jorge A. Vila jv84@cornell.edu
https://orcid.org/0000-0001-7557-9350

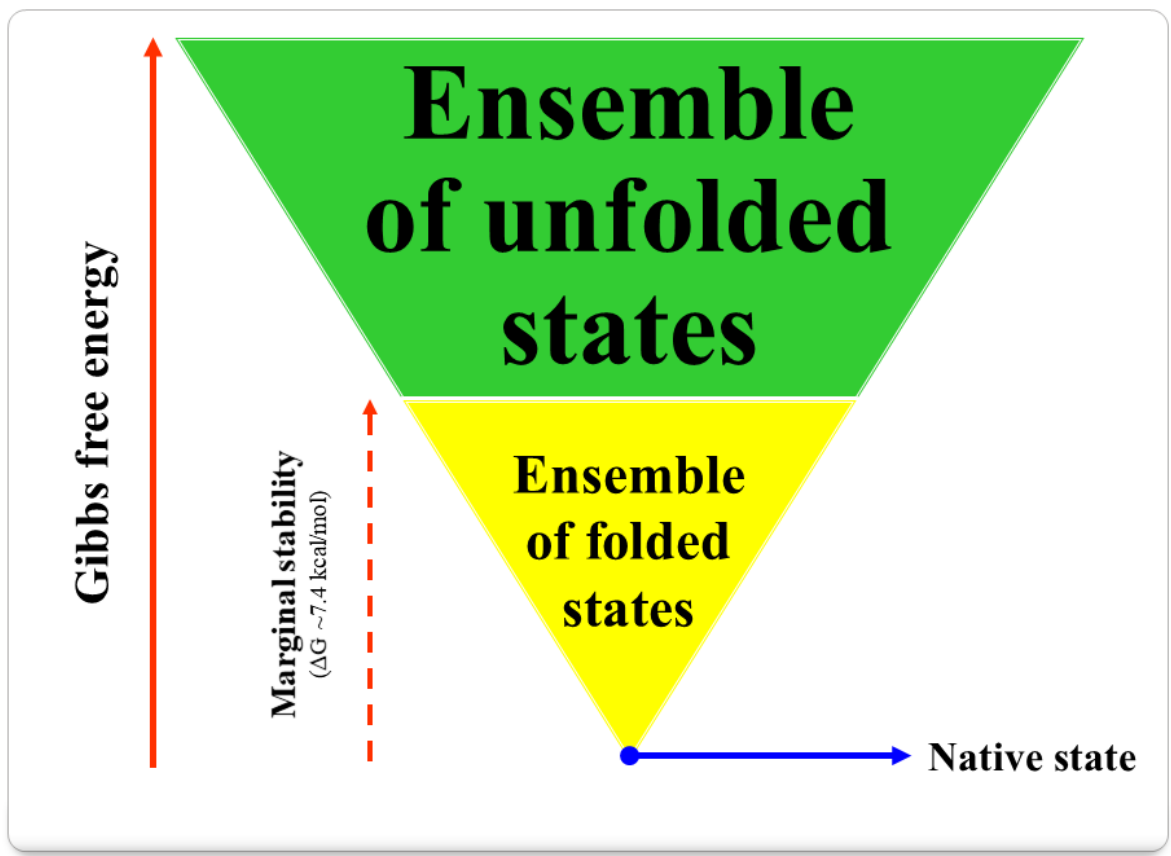

**Figure 1**. The shown schematic representation of the proteins conformational space aims to highlight the relative free-energy difference between the ensemble of folded states (yellow zone) and that of the unfolded states (green zone). The relative size of both ensembles is far from real and have only an illustrative purpose. The 'unique three-dimensional structure' representing the native state, according to Anfinsen's dogma, is highlighted in blue.